\documentclass[rnote]{aa} 

\usepackage{graphicx}
\usepackage{txfonts}
\usepackage{natbib}
\usepackage{amssymb}

\newcommand{\MLd}{\Upsilon_{\rm{d}}}
\newcommand{\kms}{\ km\ s$^{-1}$}
\newcommand{\Om}{\Omega_0}
\newcommand{\Omb}{\Omega_{{\rm b},0}}
\newcommand{\Oml}{\Omega_{\Lambda,0}}
\newcommand{\fb}{f_{{\rm b},0}}
\newcommand{\lsun}{\;L_\odot}
\newcommand{\msun}{\;M_\odot}
\newcommand{\Rvir}{R_{\mathrm{vir}}}
\newcommand{\Vmax}{V_{\rm max}}
\newcommand{\Vrot}{V}
\newcommand{\Vvir}{V_{\mathrm{vir}}}
\newcommand{\Mvir}{M_{\mathrm{vir}}}
\newcommand{\Jvir}{J_{\mathrm{vir}}}
\newcommand{\Dvir}{\Delta_{\mathrm{vir}}}
\newcommand{\rhovir}{\overline{\rho}_{\mathrm{vir}}}
\newcommand{\md}{m_{\rm{d}}}
\newcommand{\jd}{j_{\rm{d}}}
\newcommand{\id}{\iota_{\rm{d}}}
\newcommand{\ld}{\lambda_{\rm{d}}}
\newcommand{\Rd}{R_{\rm{d}}}
\newcommand{\Md}{M_{\rm{d}}}
\newcommand{\Jd}{J_{\rm{d}}}
\newcommand{\Vd}{V_{\rm{d}}}
\newcommand{\Mh}{M}

\newcommand{\fc}{f_c}
\newcommand{\fRd}{f_{R\rm{d}}}

\newcommand{\fVm}{f_{\rm vmax}}
\newcommand{\hi}{{H$\,$\footnotesize I}}

\begin{document}

\title{Fitting functions for a disk-galaxy model with different $\Lambda$CDM-halo profiles}

\titlerunning{Fitting functions for a disk-galaxy model}

   \subtitle{}

   \author{L.\ Darriba
          \and
          J.M.\ Solanes
          }

   \institute{Departament d'Astronomia i Meteorologia and Institut de Ci\`encies del Cosmos, Universitat de Barcelona. C/ Mart\'{\i} i Franqu\`es, 1; E--08028~Barcelona, Spain\\
              \email{ldarriba@am.ub.es, jm.solanes@ub.edu}
             }

   \date{Accepted: 7 April 2010}


  \abstract
  {} 
  {We present an adaptation of the standard scenario of disk-galaxy
    formation to the concordant $\Lambda$CDM cosmology aimed to derive 
    analytical expressions for the scale length and rotation
    speed of present-day disks that form within four different,
    cosmologically motivated protogalactic dark matter halo-density
    profiles.}
  {We invoke a standard galaxy-formation model that includes virial
    equilibrium of spherical dark halos, specific angular momentum
    conservation during gas cooling, and adiabatic halo response to
    the gas inflow. The mean mass-fraction and mass-to-light ratio of
    the central stellar disk are treated as free parameters whose
    values are tuned to match the zero points of the observed
    size-luminosity and circular speed-luminosity relations of
    galaxies.}
  {We supply analytical formulas for the characteristic size and
    rotation speed of disks built inside Einasto $r^{1/6}$, Hernquist,
    Burkert, and Navarro-Frenk-White dark matter halos. These
    expressions match simultaneously the observed zero points and
    slopes of the different correlations that can be built in the
    $RV\!L$ space of disk galaxies from plausible values of the galaxy-
    and star-formation efficiencies.}
   {}


   \maketitle
%

\section{Introduction}

In the current hierarchical galaxy-formation paradigm disk-galaxies
are born out of the hot gas-atmospheres associated with the potential
well of virialized cold dark matter (CDM) halos. It is assumed that
baryons have initially both the same density profile and specific
angular momentum distribution as DM -- the latter achieved, for
instance, through tidal interactions with neighboring objects in the
precollapse phase \citep[e.g.][]{Pee69}. As the gas radiates its
energy it cools and starts to fall towards the center of the DM halo 
maintaining its specific angular momentum, where it settles into a
rotationally supported disk. The assembly of a concentration of cold
baryons at the bottom of the gravitational potential well on
timescales longer than the free-fall time produces the adiabatic
contraction of the dark halo\footnote{In modern literature, the mode
  and amount of halo contraction are actually a matter of debate
  \citep[e.g.][]{Dut07,Tis09}. The outcome, however, remains
  unchanged: the properties of disk galaxies are linked to those of
  their host halos.}. In this standard picture, the internal
properties of disk galaxies are expected to be largely dictated by
those of their host halos, and through the latter, by those of the
background cosmology too.

Theoretical predictions for the distribution of disk galaxies in the
space of disk scalelength (or size), fiducial (usually, maximum or
asymptotic) rotational speed, and luminosity (or mass) based,
partially or totally, on the scenario just outlined are abundant in
the literature \citetext{e.g., \citealt*{MMW98}, hereafter
  \citeauthor{MMW98}; \citealt{Piz05,Dut07}}. They are widely used in
semi-analytic cosmological models, pre-prepared numerical simulations
of galaxy groups and clusters, and studies of disk-galaxy scaling
relations.

While nowadays there are extensive and comprehensive investigations of
the correlations between disk-galaxy properties that deal with the
scatter and covariances of the variables and allow for different modes
of halo contraction \citep[e.g.][]{Dut07}, it is not always feasible
to implement such sophisticated treatments whenever one needs to
estimate the scaling of the basic structural and kinematic parameters
of galaxies. The simplest alternative is the use of scaling laws
derived directly from fits to a given set of observations. However,
because of their lack of theoretical foundation, these formulas cannot
be extrapolated to explain the properties of galaxies other than those
from which they are derived. Halfway between these two options is
the possibility of using analytical expressions endowed with a
physical basis that enables their application to a wide range of
galactic and halo parameters. It is precisely with this aim that we
here introduce a self-consistent pure disk-formation model that
follows the well-known approach by \citeauthor{MMW98} adapted to the
canonical $\Lambda$CDM concordance cosmology and to four different
mass-density distributions for the protogalactic dark halos. This
updated scenario is capable of matching \emph{simultaneously} with
very good accuracy the zero points and slopes of the observed
correlations in the $RV\!L$ space of disk galaxies from reasonably
realistic values of its input parameters. Yet its most valuable
characteristic is its ease of implementation, as we approximated
the model predictions for the scale length and rotation speed of disks
by analytical expressions. The supplied equations can come in handy
for situations that require the generation of large numbers of
galaxies with intrinsic attributes in good agreement with the mean
observed trends, especially when the relative abundances of these
objects are known in advance.

\section{Model components}\label{galaxy_model}

We recap here the key assumptions and associated equations of our
self-consistent $\Lambda$CDM-model of disk-galaxy formation:

\begin{enumerate}
\item \emph{In the protogalactic state, the (hot) baryons and dark matter
  are well mixed within virialized spherical halos. Both
  components have the same distribution of specific angular momentum.}
\end{enumerate}
The total angular momentum $\Jvir$ of a galactic halo of virial mass
$\Mvir$ is commonly characterized in terms of the dimensionless spin
parameter
\begin{equation}\label{spin}
\lambda=\frac{\Jvir/\Mvir}{\sqrt{2}\Rvir\Vvir}\fc^{1/2}\;,
\end{equation}
which, according to the results of $N$-body simulations, follows a
lognormal distribution with median $\lambda_0$ lying in the range
$0.03\la \lambda_0\la 0.05$ \citetext{see, for instance,
  \citealt{Sha06} and references therein}, nearly independent of
cosmology, halo environment, and redshift \citep[e.g.][]{LK99}. In
Eq.~(\ref{spin}), $\Rvir$ is the virial radius inside which the
halo mean density, $\rhovir$, is $\Dvir$ times the mean density of the
universe at the redshift of observation, $\Vvir^2=G\Mvir/\Rvir$, and
$\fc$ is a dimensionless function of the halo concentration (see
below) that measures the deviation of the protogalactic halo's energy
from that of a singular isothermal sphere with the same mass,
$-(1/2)\Mvir\Vvir^2$ \citetext{see \citeauthor{MMW98}}. For the family
of flat $(\Omega_{\rm m}+\Omega_{\Lambda}=1)$ cosmogonies,
$\Dvir(z)\simeq\{18\pi^2+82[\Omega(z)-1]-39[\Omega(z)-1]^2\}/\Omega(z)$
\citep{BN98}.

The halo concentration parameter, $c$, characterizes the overall shape
of a halo density profile by measuring the ratio between its outer
radius and inner scalelength. Originally introduced for the
Navarro-Frenk-White function, its mean values are strongly correlated
with the halo mass given a cosmology \citep*[e.g.][]{NFW97}. We
approximate the mean concentration-mass relation at $z=0$ in the range
of halo masses of interest, $10^{10}\la \Mh/(h^{-1}\msun) \la
10^{13}$, by the best-fitting power-law relation recently inferred by
\citet{Mac08} from relaxed halos simulated in the Wilkinson Microwave
Anisotropy Probe 5 years results (\emph{WMAP}5) cosmology
\begin{equation}\label{c-M}
c(\Mvir,0)=9.35\left[\frac{\Mvir}{10^{12}\,h^{-1}\msun}\right]^{-0.094}\;,
\end{equation}
where $c\equiv\Rvir/r_{-2}$ is defined here in a profile-independent
form by adopting as the inner characteristic radius of the halo
density profile the radius $r_{-2}$, at which its effective logarithmic
density slope $\gamma\equiv{\mathrm{d}}\ln\rho(r)/{\mathrm{d}}\ln
r$ equals $-2$. The \emph{WMAP}5 cosmological parameters are
$(\Om,\Oml,\Omb,h,\sigma_8,n)=(0.26,0.74,0.044,0.72,0.8,0.96)$,
implying that $\rhovir\sim 96$ times the critical density for closure
at the current epoch.

\begin{enumerate}
\setcounter{enumi}{1}
\item \emph{Disks form smoothly out of cooling flows preserving the specific
  angular momentum of the baryons. The cold gas settles in centrifugal
  equilibrium at the center of the halo's potential well following an
  exponential distribution.}
\end{enumerate}

The fraction of baryons that collect into the central galaxy (in the
form of stars + cold gas) is defined as
\begin{equation}\label{md}
\md=\Md/\Mvir\;,
\end{equation}
where the values of this parameter, for which a plausible upper limit
is the universal baryon fraction $\fb=\Omb/\Om\simeq 0.17$, do not
seem to depend much on the halo mass or spin \citep{Sal09}. Similarly,
the angular momentum of the disk, expressed in units of that of its
surrounding halo, can be written as
\begin{equation}\label{jd}
\jd=\Jd/\Jvir\;.
\end{equation} 
The common yet uncertain assumption that the specific angular momenta
of the central disk galaxy and of the halo hosting it are equal,
$\Jd/\Md=\Jvir/\Mvir$, is equivalent to setting $\jd=\md$.

On the other hand, a thin exponential mass distribution of total mass
$\Md$, surface density $\Sigma(R)=\Md/(2\pi\Rd^{2})\exp(-R/\Rd)$,
and a rotation curve $\Vrot(R)$, has a total angular momentum
\begin{equation}\label{Jd}
\Jd=2\Md\Rd\Vvir\frac{1}{2}\int_0^{\infty}u^2\mbox{e}^{-u}\frac{\Vrot(u\Rd)}{\Vvir}{\mathrm{d}}u\equiv2\Md\Rd\Vvir\fRd^{-1}\;,
\end{equation}
where the factor $\fRd$ is unity for a disk with a flat rotation curve
at the level $\Vvir$, and where the total circular speed is computed
by summing in quadrature the contributions from the disk of cold
baryons, $\Vd$, and from the dark halo, $V_{\rm{h}}$,
\begin{equation}\label{vrot2}
\Vrot^2(R)=V_{\rm{d}}^2(R)+V_{\rm{h}}^2(r)_{r=R}\;,
\end{equation}
with $R$ the cylindrical radius. An expression for $V_{\rm{d}}^2(R)$
can be found in \citet{BT08}, p.101, Eq.~(2.165).

Substituting Eqs.~(\ref{spin}), (\ref{md}), and (\ref{jd}) into
Eq.~(\ref{Jd}), one can then obtain the disk scalelength as a function
of the model parameters
\begin{equation}\label{scalelength}
\Rd=\frac{\fRd}{\sqrt{2\fc}}\ld\Rvir\;,
\end{equation}
with $\ld\equiv(\jd/\md)\lambda$ the effective spin of the disk.

\begin{enumerate}
\setcounter{enumi}{2}
\item \emph{The halo contracts adiabatically and without shell crossing to
  gas inflow.}
\end{enumerate}

According to the adiabatic compression paradigm \citep{Blu86}, for a
spherical halo in which all particles move on circular orbits with
velocity $V(r)$, any function of the specific angular momentum $rV(r)$
is an adiabatic invariant. Assuming that the initial and final mass
distributions of the different components are spherically symmetric, 
this invariance leads to the relationship
\begin{equation}
  [\Md(r)+M_i(r_i)(1-\md)]r=M_i(r_i)r_i\;,
\end{equation}
where $r_i$ and $r$ are, respectively, the initial and final radius of
the spherical shells, $M_i(r)$ is the initial protogalactic halo mass
profile, and $\Md(r)$ comes from the replacement of the final thin
exponential disk configuration by the spherical density profile that
has the same enclosed mass.

The contribution to the total rotation curve (Eq.~[\ref{vrot2}]) from
the dark matter (and the remaining hot baryons) is therefore
\begin{equation}\label{vroth2}
V_{\rm{h}}^2(r)=GM_i(r_i)(1-\md)/r\;.
\end{equation}

Taking into account that both halo and disk properties are directly
proportional to their corresponding virial parameters,
Eq.~(\ref{vrot2}) allows one to express the amplitude of the total
rotation curve at a given number of scalelengths and, in particular,
its peak value, $\Vmax$, in the compact form (cf.\ \citeauthor{MMW98})
\begin{equation}\label{vrots}
\Vmax=\Vvir\fVm\;,
\end{equation}
with $\fVm$ a dimensionless factor that, like $\fRd$, depends on the
adopted halo density law and on the values of parameters $\ld$, $c$,
and $\md$.

\section{Model predictions}\label{model_predictions}

We now proceed to tune the free parameters of our disk-galaxy
formation model to match the scaling relations in $RV\!L$
space observed at $z\sim 0$. For a given halo virial mass, two are the
free parameters in our modeling: the disk mass fraction, $\md$, and
mass-to-light ratio, $\MLd\equiv\Md/L$. This latter quantity is needed
to convert the predicted disk masses into observed luminosities. We do
not allow the average effective disk spin to vary freely however,
but use the condition $\jd=\md$ to set it equal to three
representative values of $\lambda_0$: 0.03, 0.04, and 0.05\footnote{We
  ignore here a possible dependence of this parameter on halo mass
  \citep[e.g.][]{Ber08}.}.

\begin{table}
\caption[]{Halo profiles.}
\label{profiles}
\centering
\renewcommand{\footnoterule}{}
\begin{tabular}{lccl} 
\hline
\noalign{\smallskip}
Profile &  $\tilde{\rho}(s)$ & $c/c_{\rm{s}}$ & \multicolumn{1}{c}{Reference} \\
\hline
\noalign{\smallskip}
        \citeauthor{EH89}$_{\rm{6}}$ & $\exp(1-s^{1/6})$ & $12^{-6}$ & \citet*{EH89}\\
        \citeauthor{Her90} & $8/s/(1+s)^3$ & 2 & \citet*{Her90}\\
        \citeauthor{Bur95} & $4/(1+s)/(1+s^2)$ & 0.6573 & \citet*{Bur95}\\
        \citeauthor{NFW97} & $4/s/(1+s)^2$ & 1 & \citet*{NFW97}\\
\hline
\end{tabular}
\end{table}

We investigated the performance of our model for the
four functional forms of protogalactic DM halos listed in
Table~\ref{profiles}. They are among the most representative functions
used in the literature to describe the equilibrium density profiles of
halos generated in CDM $N$-body simulations. All of them are spherical
density distributions of the form
\begin{equation}
\rho(r)=\rho_{\rm{s}}\tilde{\rho}(s)\;,
\end{equation}
where $\rho_{\rm{s}}\equiv\rho(r_{\rm{s}})$ and $r_{\rm{s}}$ are the 
characteristic density and scale radius of the profile respectively, 
and $\tilde{\rho}(s)$ is a dimensionless function of the dimensionless 
radius $s\equiv r/r_{\rm{s}}$.

With the aid of the $c(\Mvir)$ relation these expressions can be
reduced to uniparametric\footnote{The Einasto $r^{1/n}$ model has an
  additional parameter $n$ controlling the curvature of the
  profile. In our modeling this parameter is kept fixed to $n=6$, a
  value representative of galaxy-sized halos \citep{Mer05}.} density
laws in which the halo structure is fully determined from $\Mvir$. It
can be shown that
\begin{equation}
\rho_{\rm{s}}=\frac{1}{3}\,\rhovir c_{\rm{s}}^3\int_0^{c_{\rm{s}}}s^2\tilde{\rho}(s){\mathrm{d}}s\;,
\end{equation}
where the characteristic concentration
$c_{\rm{s}}\equiv\Rvir/r_{\rm{s}}$ is directly related to the
profile-independent halo concentration parameter $c$ defined in
Eq.~(\ref{c-M}) (see Table~\ref{profiles}).

In order to constrain our model predictions, we consider a subset of
the SFI++ sample \citep{Spr07} consisting of 649 galaxies also
included in the compilation of $\sim 1300$ local field and cluster
spiral galaxies by \citet{Cou07}. The full SFI++ contains measures of
intrinsic rotation velocity widths reduced to a homogeneous system
based on the 21 cm spectral line, $W$, as well as absolute $I$-band
magnitudes for near 5000 spiral galaxies, while the dataset by
\citeauthor{Cou07} provides inclination-corrected estimates of disk
scalelengths also in the $I$-band (below both observables
and model parameters will refer to this near-IR band).

As stated by \citet{Cat07}, for most intermediate and bright disks
the width of the global \hi\ profile provides a more reliable
observational estimate of the peak rotation velocity than the widths
of H$\alpha$ rotation curves, at least for objects not affected by
environmental interactions. This is probably because the latter are
usually evaluated either at a radius where, on average, they are still
rising (e.g., $2.2\Rd$), or on the asymptotic part of the optical
disk. Accordingly, we adopt the approximation $W/2\simeq\Vmax$, where
$\Vmax$ is the maximum width of our model total speed curve measured
within $5\Rd$.

\subsection{Scaling laws}

The distribution of $R$ as a function of $V$ provides the most
effective way of determining the value of $\md$ -- which for bright
galaxies represents to a good approximation the stellar mass fraction
-- that best fits the observations for each one of the values of $\ld$
under consideration. To allow for a more robust comparison between the
model predictions and the data, the $RV$ scaling law has been recast
in the form of the tighter relation between the average specific
angular momentum of disks computed from the fiducial rotation speed of
the galaxies, $\id=2\Rd\Vmax$, and $\Vmax$. In a log-log scale this
relationship is expected to follow a straight line with a slope near 2
and a zero point that is a sensitive function of $\md$.

In the upper-left panel of Fig.~\ref{figallSB}, we show the model
relations that best fit the barycenter of the data cloud for the four
halo profiles considered and the central value of $\ld$ (the best
values of $\md$ obtained for each one of the three values adopted for
$\ld$ are listed in Col. 3 of Table~\ref{parameters}). It can be
seen from this plot that our disk models also reproduce the slope of
the observed $\id-\Vmax$ scaling law. We note in passing that on the
basis of its location in this diagram, the angular momentum and disk
scale of the Milky Way (MW) are unrepresentative of those of a typical
spiral \citep[see also][]{Ham07}.

\begin{table}
\caption[]{Model parameters.}
\label{parameters}
\centering
\begin{tabular}{lcccc} 
\hline
\noalign{\smallskip}
Profile      &  $\ld$ & $\md$ &  $\MLd/[h(M/L_I)_\odot]$ & $\Vmax/\Vvir$ \\
\hline
\noalign{\smallskip}
            \citeauthor{EH89}$_{\rm{6}}$ & 0.03 & 0.025 & 1.20 & 1.4  \\
            \citeauthor{EH89}$_{\rm{6}}$ & 0.04 & 0.050 & 1.50 & 1.6  \\
            \citeauthor{EH89}$_{\rm{6}}$ & 0.05 & 0.080 & 1.70 & 1.8  \\
            \citeauthor{Her90} & 0.03 & 0.020 & 1.00 & 1.4  \\
            \citeauthor{Her90} & 0.04 & 0.035 & 1.25 & 1.5  \\
            \citeauthor{Her90} & 0.05 & 0.055 & 1.45 & 1.7  \\
            \citeauthor{Bur95} & 0.03 & 0.030 & 1.50 & 1.4  \\
            \citeauthor{Bur95} & 0.04 & 0.050 & 1.60 & 1.6  \\
            \citeauthor{Bur95} & 0.05 & 0.075 & 1.75 & 1.7  \\
            \citeauthor{NFW97} & 0.03 & 0.030 & 1.30 & 1.4  \\
            \citeauthor{NFW97} & 0.04 & 0.050 & 1.50 & 1.6  \\
            \citeauthor{NFW97} & 0.05 & 0.080 & 1.70 & 1.8  \\
\hline
\end{tabular}
\end{table}

With $\md$ fixed and given that the halo concentration is not allowed
to vary freely, the most sensitive tuning of the other free parameter
of the model, $\MLd$, is achieved by normalizing the model predictions
to the observed $V\!L$ relation. For the latter, which is fully 
independent of surface brightness \citep{Zwa95,CR99}, we use the
calibration of the Tully-Fisher (TF) relationship corrected from
observational and sample biases calculated by \citet{Mas06} using 807
cluster galaxies extracted from the SFI++ catalog, which we rewrite in
the form
\begin{equation}\label{TF}
M_I-5\,\log\,h= -20.85-7.85\,[\log\,(2\Vmax)-2.5]\;
\end{equation}
to facilitate the comparison with our model predictions. In
Eq.~(\ref{TF}), $M_{I,\odot}=4.11$~mag has been adopted to transform
model luminosities into absolute magnitudes. As in the former case,
the upper-right panel of Fig.~\ref{figallSB} shows that the predicted
$V\!L$ relations (again we show only those inferred using the central
value of $\ld$) closely match the slope of the empirical estimate. 
In this case, the best values of $\MLd$ have been set by minimizing the
residual between the model predictions and the observed TF
relationship over the full available range of velocities. As could be
expected, the agreement between the model predictions and our
SFI++-based comparison sample is fairly good too. This panel also
illustrates the well-known deficiency in luminosity of the MW with
respect to the TF relation \citep{Por07}.

The excellent agreement between predictions and observations in the
$RV$ and $V\!L$ planes is maintained for the joint distribution of the
three variables. The lower-left panel of Fig.~\ref{figallSB} depicts,
again for the central value of $\ld$, the scatter diagram of central
disk surface density, $\Sigma_0=\Md/(2\pi\Rd^{2})$, and rotation speed. 
We have converted \citeauthor{Spr07}'s data on $M_I$ into total disk
luminosities, which in turn have been transformed into disk masses using 
the values of $\MLd$ derived from the normalization of the $V\!L$ relation. 
It can be seen that our model predictions are once more comfortably close 
to both the normalization and, in this case, \emph{curved} mean trend delineated by
the data.

\subsection{Fitting functions for galaxy scaling parameters}

By using the values quoted in Cols.\ 3 and 4 of Table~\ref{parameters}
it is straightforward to calculate the average luminosity of a nearby
disk embedded in a halo of given $\Mvir$ and $\ld$. However, as shown
in Sect.~\ref{galaxy_model}, each of the remaining fundamental disk
properties, the characteristic scale and rotation speed, participates
in the calculation of the other. As a result, they can only be
computed by applying an iterative procedure that, despite its fast
convergence, remains cumbersome. For this reason, it is very
convenient to approximate the dimensionless factors $\fRd$ and $\fVm$
appearing in the calculation of $\Rd$ and $\Vmax$
(eqs.~[\ref{scalelength}] and [\ref{vrots}], respectively) by fitting
functions. Drawing inspiration from \citeauthor{MMW98}, we propose the
following fitting formulas, which are valid for any of the
protogalactic halo mass density profiles explored:
\begin{eqnarray}\label{fRfV}
\lefteqn{\fRd(\ld,c,\md) \approx \left(\frac{\ld}{0.05}\right)^{a_1+a_2\md/(1+\ld)}(a_3+a_4\md+a_5\md^2)}\hspace{.64 in}\nonumber \\
& & \times\;[a_6+a_7c+a_8/c+a_{9}/(c\md)]\;,\\
\lefteqn{\fVm(\ld,c,\md) \approx \left(\frac{\ld}{0.05}\right)^{0.001/\md+b_1\md}(b_2+b_3\md+b_4\md^2)}\hspace{.65 in}\nonumber \\
& & \times\;(b_5+b_6c+b_7/c)\;.
\end{eqnarray}
The values of the coefficients corresponding to each profile, which
are independent of the adopted $c(\Mvir)$ relation, are listed in
Table~\ref{coefficients}. Both approximations are accurate to within 
8 $\%$ for $5\leq c\leq 25$, $0.03\leq\ld\leq 0.05$, and
$0.01\leq\md\leq 0.1$. We note that attempts to fit these
dimensionless factors using either solely polynomials or a linear
combination of power laws of parameters $\ld$,
$c$, and $\md$ have required a substantially larger number of terms to
achieve a similarly satisfactory match. Given the countless number of
real functions that can be implemented, it would be obviously possible
to find other formulae also producing good fits, but most likely they
will be more complicated than the above expressions.

\begin{table*}
\caption[]{Coefficients of the approximations.}
\label{coefficients}
\centering
\begin{tabular}{lccccccccc} 
\hline
\noalign{\smallskip}
Profile      &  $a_1$ & $a_2$ & $a_3$ & $a_4$ & $a_5$ & $a_6$ & $a_7$ & $a_8$ & $a_9$ \\
\noalign{\smallskip}
\hline
\noalign{\smallskip}
            \citeauthor{EH89}$_{\rm{6}}$ & $-$0.041 & 5.95 & 0.264 & $-$1.45 & 3.35 & 3.16 & $-$0.0311 & 5.58 & $-$  \\
            \citeauthor{Her90} & $-$0.037 & 4.86 & 0.252 & $-$1.73 & 6.14 & 3.13 & $-$0.0429 & 7.62 & $-$  \\
            \citeauthor{Bur95} & $-$0.058 & 5.50 & 0.275 & $-$1.20 & 0.70 & 2.63 & $-$0.0234 & 7.44 & 0.027  \\
            \citeauthor{NFW97} & $-$0.056 & 6.18 & 0.267 & $-$1.46 & 3.31 & 3.19 & $-$0.0310 & 5.56 & $-$  \\
\hline
\noalign{\smallskip}
                  &  $b_1$ & $b_2$ & $b_3$ & $b_4$ & $b_5$ & $b_6$ & $b_7$ \\
\noalign{\smallskip}
\hline
\noalign{\smallskip}
            \citeauthor{EH89}$_{\rm{6}}$ & $-$5.50 & 0.634 & 5.24 & 0.78 & 1.44 & 0.0343 & $-$0.72 \\
            \citeauthor{Her90} & $-$5.58 & 0.792 & 6.78 & 0.16 & 1.35 & 0.0351 & $-$1.77 \\
            \citeauthor{Bur95} & $-$4.24 & 0.722 & 4.91 & 2.24 & 1.32 & 0.0394 & $-$1.03  \\
            \citeauthor{NFW97} & $-$5.38 & 0.651 & 5.38 & 0.84 & 1.41 & 0.0346 & $-$0.82  \\
\hline
\end{tabular}
\end{table*}

\section{Discussion and conclusions}

We formulated a standard formation model of disk galaxies inside
DM halos within the concordant $\Lambda$CDM cosmology that
simultaneously predicts with high accuracy the main trends of the
observed fundamental scaling relations of nearby galaxies in $RV\!L$
space. This modeling has been developed with the sole aim of deriving
physically sound analytical expressions for predicting the central
properties that characterize the light profiles and rotation curves of
\emph{typical spirals}. We supply formulas for Einasto $r^{1/6}$, Hernquist,
Burkert, and Navarro-Frenk-White protogalactic halo mass density
distributions that provide a similarly good overall description of the
data on disks for realistic enough values of the model free
parameters. We find that, for a given $\ld$, the predictions of the
Einasto $r^{1/6}$, Hernquist, and Navarro-Frenk-White profiles are
relatively similar, while the Hernquist profile -- the only density
law investigated that does not follow a $\rho\propto r^{-3}$ behavior
near $\Rvir$ -- requires values of $\md$ and $\MLd$ that are lower by
about a factor of 0.70 and 0.85, respectively.

\begin{figure*}
\centering
  \begin{center}
    \begin{tabular}{cc}
      \resizebox{74mm}{!}{\includegraphics{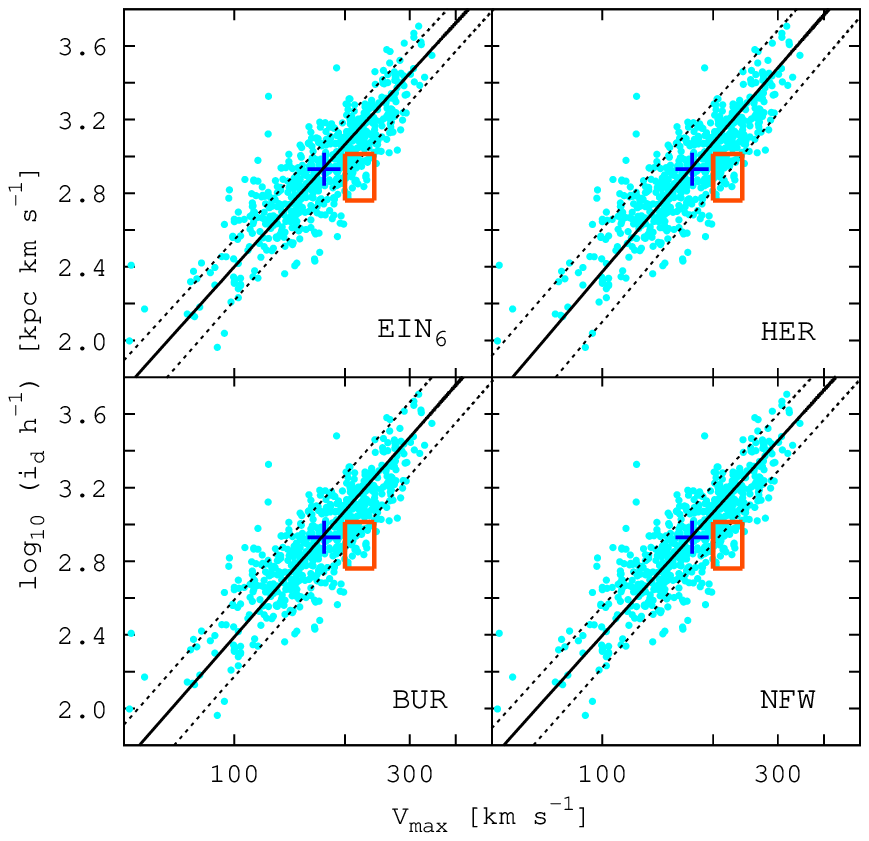}} &
      \resizebox{74mm}{!}{\includegraphics{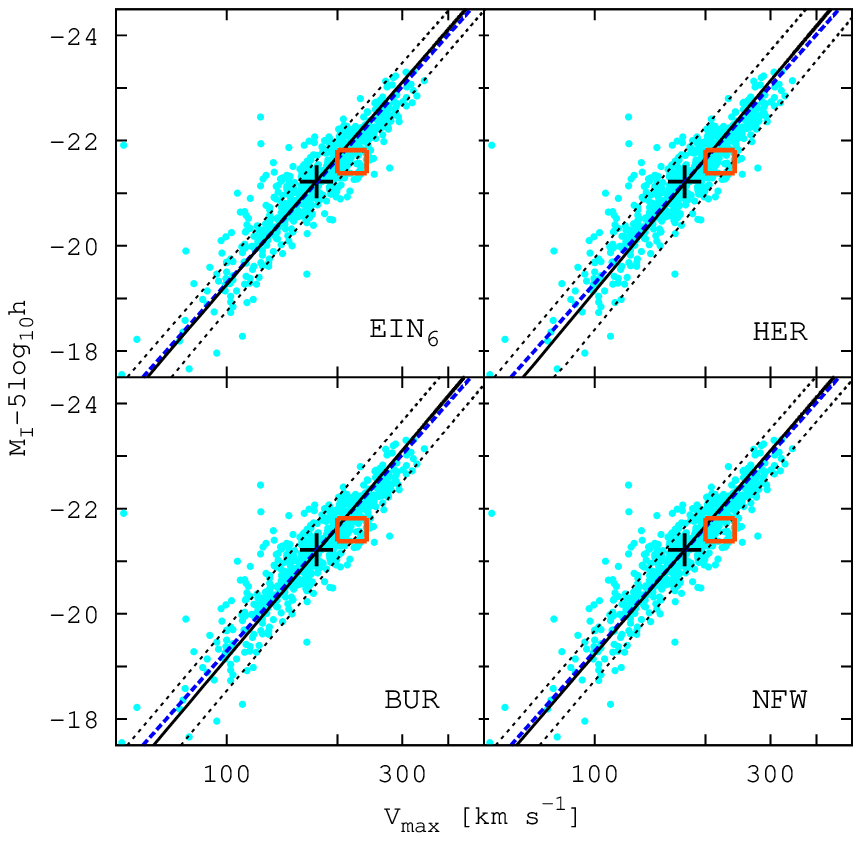}} \\
      \resizebox{74mm}{!}{\includegraphics{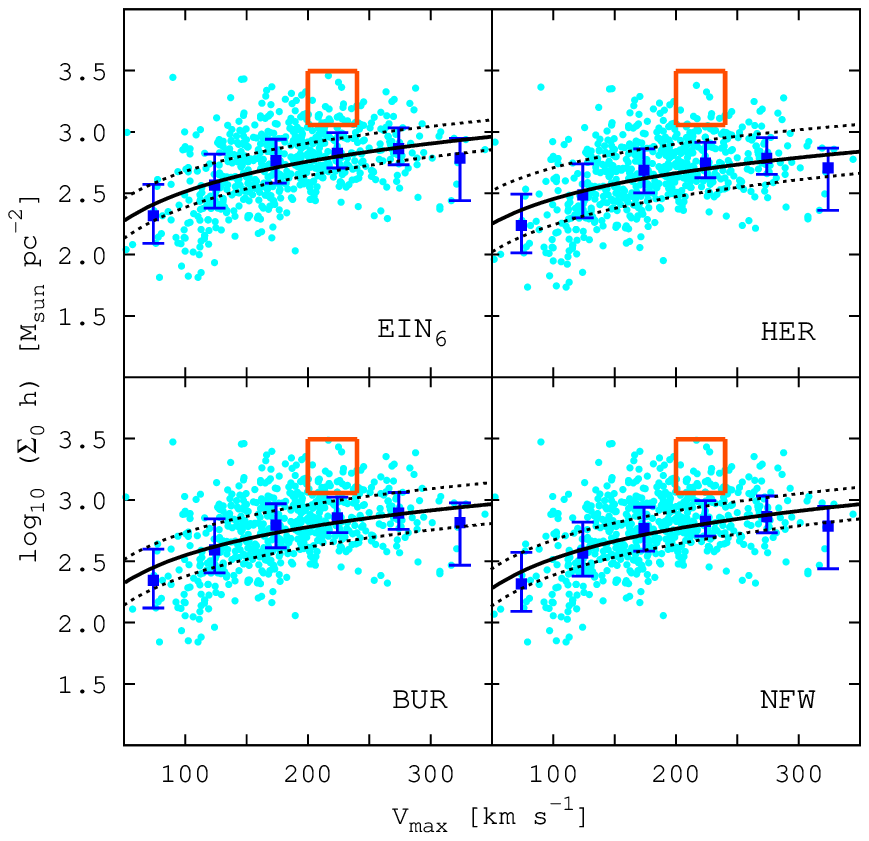}} &
    \end{tabular}

    \caption{Scale relations for nearby disks. \emph{Upper-left:}
      Disk-specific angular momentum as a function of $\Vmax$. The
      cross shows the barycenter of the data. \emph{Upper-right:}
      $I$-band TF relation. The dashed line shows the one derived by
      \citet{Mas06} from SFI++ data \citep{Spr07}. The cross 
      shows the barycenter of the data. \emph{Lower-left:} Central-disk 
      surface-mass density vs.\ $\Vmax$. Squares with
      error bars show the median observational values and the first
      and third quartiles in each velocity bin. In all these plots the
      solid lines show model predictions for pure disks with
      $\ld=0.04$ using the mean $c(\Mvir)$ relation at $z=0$, while
      the dotted curves show from top to bottom the predictions
      resulting from adopting the 2.3th and 97.7th percentiles of the
      concentration distribution ($2\sigma$ scatter) assuming
      $\sigma_{\log c}=0.11$ \citep{Mac08}. The error boxes represent,
      for comparison, the values measured for the MW from: $\Rd=2.5\pm
      0.5$ kpc, $\Md=5\pm 0.5\times 10^{10}\msun$, and $\Vmax=220\pm
      20$\kms\ \citep{BT08}, and $L_I=(4\pm 0.8)\times 10^{10}\lsun$
      \citep{Por07}. The data clouds are build on the compilation of
      $I$-band absolute magnitudes and \hi\ rotation widths by
      \citet{Spr07} and on the $I$-band disk scalelength measurements
      by \citet{Cou07}. \label{figallSB}}
  \end{center}
\end{figure*}

The reader may have noticed that our best models yield for $\MLd$,
i.e. for the inverse of the average star-formation efficiency, values
somewhat lower than those inferred from population synthesis
calculations \citep[e.g.][]{Piz05}. We stress however that
the observational estimates of this parameter are affected by
considerable uncertainties, our prediction that the average
mass-to-light ratio of disks is $\sim 1 (M/L_I)_\odot$, which is 
consistent with submaximal disks arguments \citep{CR99,KdN08}, as well
as relatively close to the values adopted as input in more
sophisticated models of disk formation \citep{Dut07}. On the other
hand, we find that the predicted values of $\md$ are directly
correlated with those adopted for $\ld$. In particular we note that a
value of $\ld=0.03$, which coincides with the median of the
distribution of the spin parameter for relaxed halos derived by
\citet{Mac08}, implies a small current average galaxy-formation
efficiency, $\md/\fb<0.2$. This agrees well with the predictions
of galaxy evolution from halo occupation models \citep{ZCZ07} and
methods that match the stellar mass function to that of the halo 
\citep{CW09}. Further recent support for low $\md$ (and $\ld$,
according to our model) comes for instance from weak lensing
measurements \citep{Man06} and from the roughly universal
distributions of this parameter obtained by \citet{Sal09} for various
implementations of feedback in large cosmological
$N$-body/gasdynamical simulations. Notice also the fifth column in
Table~\ref{parameters}, where we list the ratio $\Vmax/\Vvir$
calculated for a MW-mass halo, which increases with increasing $\ld$
and decreasing $\Mvir$. As stated by \citet{Dut07}, the relatively
high values we predict for this ratio -- a characteristic common to
standard models -- would likely hamper a simultaneous match to the
galaxy LF that, according to semi-analytical models of galaxy
formation, requires the condition $\Vmax\sim\Vvir$.

We made no attempt to explore the scatter of the
observed scaling relations and the covariance that exists between
model parameters, except for Fig.~\ref{figallSB}, where we carry out a naive
comparison between the spread of the data and that resulting from
taking into account the predicted scale of the probability
distribution of the halo concentration. Including this and other
sources of scatter, such as the variance of the halo spin parameter,
or the dependence of the concentration-mass relation on the adopted
cosmology \citep[e.g.][]{Mac08}, would undoubtedly enrich the
analysis. Yet, a thorough investigation of scatter requires dealing
with the joint probability distribution of all the parameters entering
the model and, in particular, with all their covariances (not just the
variances), which ideally should be corrected from measurement
errors. This far exceeds the scope of our present research. 
We note in addition that efforts in the direction
just outlined will soon be much more effective when they can be
applied to objective, homogeneous, and complete $RV\!L$ datasets free
of nontrivial selection biases, as those build from the
cross-correlation of wide-area spectroscopic optical and \hi\ surveys
(e.g., Toribio et al.\ 2010, in preparation).

Finally, we wish to comment on the possibility of extending our model
predictions to distant galaxies by adopting a $c(\Mvir)$ relationship
of the form
\begin{equation}\label{c-M(z)}
c(\Mvir,z)=c(\Mvir,0)g(z)^{-1}\;,
\end{equation}
with $c(\Mvir,0)$ given in Eq.~(\ref{c-M}) and $g(z)$ the
concentration growth factor, which can be taken proportional to
$H(z)^{2/3}$, as found in a recent modification of the original
\citet{Bul01} model for \emph{WMAP} cosmologies by \citet{Mac08}.

\begin{acknowledgements}

  We thank the anonymous referee for his/her thorough review and
  appreciate the comments and suggestions, which significantly helped
  to improving the manuscript. This work is supported by the Spanish
  Direcci\'on General de Investigaci\'on Cient\'{\i}fica y T\'ecnica,
  under contract AYA2007-60366.

\end{acknowledgements}

\end{document}